# GPU System Calls

*Ján Veselý[§‡], Arkaprava Basu[‡], Abhishek Bhattacharjee[§], Gabriel H. Loh[‡], Mark Oskin[‡], Steven K. Reinhardt[‡]*

[§] *Rutgers University,* [‡] *AMD Research*

## Abstract

GPUs are becoming first-class compute citizens and are being tasked to perform increasingly complex work. Modern GPUs increasingly support programmability-enhancing features such as shared virtual memory and hardware cache coherence, enabling them to run a wider variety of programs. But a key aspect of general-purpose programming where GPUs are still found lacking is the ability to invoke system calls.

We explore how to directly invoke generic system calls in GPU programs. We examine how system calls should be meshed with prevailing GPGPU programming models, where thousands of threads are organized in a hierarchy of execution groups: Should a system call be invoked at the individual GPU task, or at different execution group levels? What are reasonable ordering semantics for GPU system calls across these hierarchy of execution groups? To study these questions, we implemented GENESYS – a mechanism to allow GPU programs to invoke system calls in the Linux operating system. Numerous subtle changes to Linux were necessary, as the existing kernel assumes that only CPUs invoke system calls. We analyze the performance of GENESYS using micro-benchmarks and three applications that exercise the filesystem, networking, and memory allocation subsystems of the kernel. We conclude by analyzing the suitability of all of Linux's system calls for the GPU.

## 1 Introduction

Graphics processing units (GPUs) are now widely used for high-performance computing (HPC), machine learning, and data-analytics. Increasing deployments of these general-purpose GPUs (GPGPUs) have been enabled, in part, by improvements in their programmability. GPUs have gradually evolved from fixed function 3D accelerators to fully programmable units [17, 25, 39]. Today, they support programmability-enhancing features such as address translation [20, 5, 22, 29, 31, 41], virtual memory [41], and cache coherence [1, 7, 14, 30, 33].

GPU programming models have steadily evolved with these hardware changes. Early GPGPU programmers [40, 23, 24, 32, 35] adapted graphics-oriented programming models such as OpenGL [10] and DirectX [16] for general-purpose usage. Since then, GPGPU programming has become increasingly accessible to traditional CPU programmers, with graphics APIs giving way to computation-oriented languages with a familiar C/C++ heritage such as OpenCL™ [9], C++AMP [15], and CUDA [19]. But access to privileged OS services, or system calls, are an important aspect of CPU programming that remain out of reach for typical GPU programmers.

While traditionally considered to only be of academic curiosity, GPU system calls have recently begun attracting wider attention [13, 21, 37, 26]. Studies have explored the first few types of system calls on GPU programs for file input/out (I/O) [37] and network access [13]. These studies present an important start but focus on specific OS subsystems, such as management of the file system buffer cache, and in some cases, rely on specific hardware features such as direct user-mode access to the network interface controller [21]. As such, they point to the exigent need for a broader examination of how system calls can aid GPU programs.

Consequently, this paper generalizes the means by which GPU programs directly invoke system calls, and the benefits of doing so. Our implementation of GPU system calls reveals two main benefits – better programmability and the potential for improved performance. In assessing how to achieve these, we study several research questions unique to GPU system calls:

**How does the GPU execution model impact system call invocation strategies?** To manage parallelism, GPU programming languages and the underlying hardware architecture decompose work into a hierarchy of execution groups. The granularity of these groups range from work-items (or GPU threads) to work-groups (composed of hundreds of work-items) to kernels (composed of hundreds work-groups) [1]. This naturally presents the following research question – at which of these granularities should GPU system calls be invoked?

**How should GPU system calls be ordered?** Traditional CPU system calls are implemented with the guar-antee that instructions prior to the system call have completed execution, while code following the system call remains unexecuted. These semantics are a good fit for CPUs, which generally target single-threaded execution. But such "strong ordering" semantics may be overly-conservative for GPUs, acting as implicit synchronization barriers across potentially thousands of work-items, compromising performance. In fact, we find that GPU code is often amenable to "relaxed ordering", where such implicit barriers may be avoided without harming the semantics of common GPU programs. We describe the situations where strong versus weak ordering decisions affect performance. Beyond ordering, we study the utility of "blocking" versus "non-blocking" GPU system calls.

[1] Without loss of generality, we use the AMD terminology of work-items, work-groups, kernels, and compute unit (CU), although our work applies equally to the NVIDIA terminology of threads, thread-blocks, kernels, and streaming multiprocessor (SM) respectively.

**Where and how should GPU system calls be processed?** Modern GPUs cannot, as of yet, execute privileged instructions or support security rings. This is unlikely to change in the near future, due to concerns about how privileged modes on GPUs may compromise security [22]. We assume, therefore, that system calls invoked by GPU programs need to ultimately be serviced by the CPU. This makes efficient GPU-CPU communication and CPU-side processing fundamental to GPU system calls. We find that careful use of modern GPU features like shared virtual addressing [41] and page fault support [29, 31], coupled with traditional interrupt mechanisms, can enable efficient CPU-GPU communication of system call requests and responses.

To study these broad research questions, we implement and evaluate a framework for **generic system call invocation** on GPUs, or **GENESYS**, in the Linux OS. We use microbenchmarks to study the questions posed above. We then showcase the benefits of GPU system calls on with three workloads: string-matching with storage I/O [37], an echo server with network I/O, and an enhanced version of a GPGPU benchmark [41] that provides hints to the memory allocator within the OS kernel.

When designing GENESYS, we ran into several interesting questions. For example, what system calls make sense to provide to GPUs? System calls such as *pread-/pwrite* to file(s) or *send/recv* to and from the network stack are useful since they underpin many I/O activities required to complete a task. But system calls such as *fork* and *execv* do not, for now, seem necessary for GPU threads. In the middle are many system calls which need adaptation for GPU execution. For example, *getrusage* can be adapted to return information about GPU resource usage, and *sched_setaffinity* likely makes sense only if the GPU thread scheduler is software-programmable, which it currently is not. We conclude our study by examining all system calls in Linux and classifying their usefulness for GPU applications.

## 2 Motivation

A reasonable research question to ponder is, why equip GPUs with system call invocation capabilities at all? After all, the lack of system call support has not hindered widespread GPU adoption.

Consider Figure 1(left), which depicts how GPU programs currently invoke system calls. At a high-level, programmers are forced to delay system call requests until the end of the GPU kernel invocation. This is not ideal because developers have to take what was a single conceptual GPU kernel and partition it into two – one before the system call, and one after it. This model, which is akin to continuations, is notoriously difficult to program for [6]. Compiler technologies can assist the process [36], but the effect of ending the GPU kernel, and restarting another is the same as a barrier synchronization across all GPU threads and adds unnecessary roundtrips between the CPU and the GPU. This significantly degrades performance.

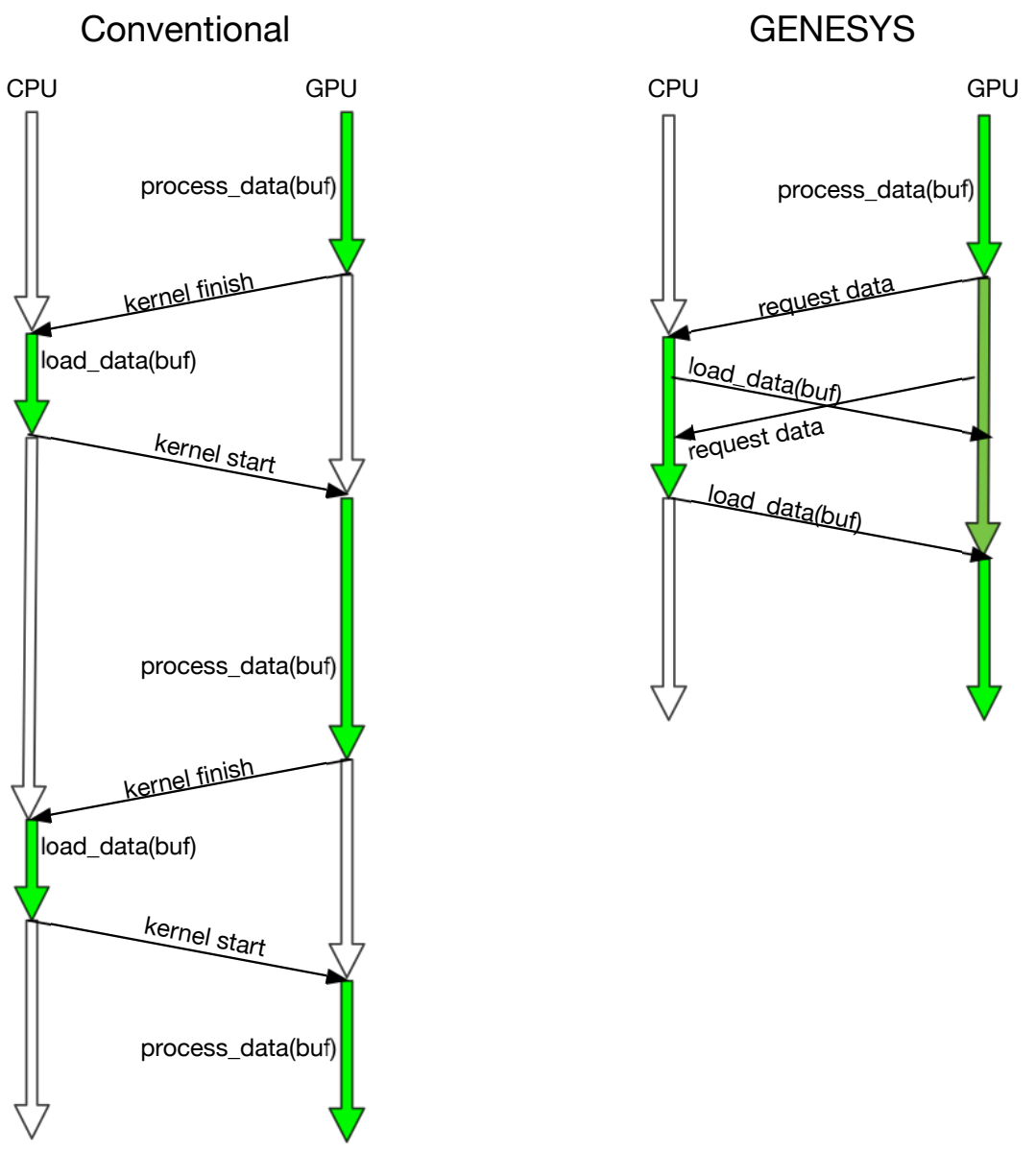

Figure 1: (Left) Timeline of events when the GPU has to rely on a CPU to handle system services; and (Right) when the GPU has system call invocation capabilities.

Now consider Figure 1(right), which shows a timeline for a system where the GPU can directly invoke the OS. There is no need for repeated kernel launches, enabling better GPU efficiency because processing can proceed uninterrupted. System calls (e.g., request_data) are still processed by the CPU, as often-times the work requires access to hardware resources only the CPU can interact with. But CPUs do not need to actively "baby-sit" the GPU and can schedule tasks in response to GPU system calls as and when needed. CPU system call processing also naturally overlaps with the execution of other GPU threads.

OSes have traditionally provided a standardized abstract machine, in the form of a process, to user programs executing on the CPU. Parts of this process abstraction, such as memory layout, the location of program arguments, an ISA, have benefited GPU programs as well. Other aspects, however, such as standardized and protected access to the filesystem, network, and memory allocator are extremely important for processes, yet lacking for GPU code. Allowing GPU code to invoke system calls is a further step to providing a more complete process abstraction to GPU code.

## 3 Our Approach

We focus on how to implement a framework for *generic* system calls. Many past studies, GPUfs [37], GPUnet [13] and CPU-to-GPU callbacks [26], have implemented specific types of system calls to enable research on the benefits of offloading aspects of filesystems and networking to GPUs. Our efforts are complementary and provide a single unifying mechanism for all system calls.

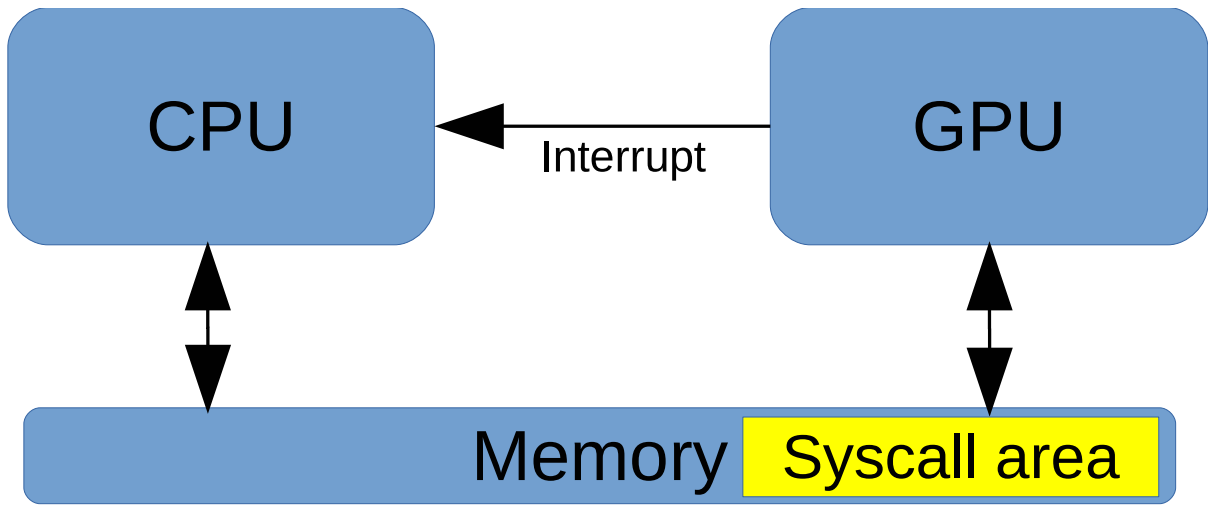

Figure 2: High-level overview of how GPU system calls are invoked and processed on CPUs.

We use the GPU system call mechanism illustrated in Figure 2 to enable our studies. We use a system with a CPU (possibly with multiple cores) and GPU. Our approach is equally applicable to integrated CPU-GPUs (also called accelerated processing units or APUs) and to discrete GPUs. We assume that both CPU and GPU can access system memory, a feature that is common today.

Figure 2 details the steps used for GPU system calls. First, the GPU invokes a system call. Since the GPU is not capable of executing system calls itself, the CPU processes them on its behalf. Therefore, the GPU must place the relevant system call arguments and information in a portion of system memory also visible by the CPU. We designate a system memory *syscall area* to store this information. In the next step, the GPU interrupts the CPU, conveying the need for system call processing. With the interrupt, the GPU also identifies the wavefront issuing the system call. This triggers the execution of an interrupt handler on the CPU, which reads the *syscall area*, processes the interrupt, and writes the results back into the *syscall area*. Finally, the CPU notifies the GPU wavefront that its system call has been processed.

Our system call invocation framework relies on the ability of the GPU to interrupt the CPU, and we do leverage readily-available hardware [2, 12] for this support. But this is not a fundamental design choice – in fact, prior work [26, 37] uses a CPU polling thread to service a limited set of GPU system service requests instead. Although using GPU-to-CPU interrupts does present certain benefits (e.g., freeing CPUs from having to poll, allowing them to execute other applications, etc.), our work is orthogonal to the interrupt versus polling debate.

Finally, while features like shared virtual memory and CPU-GPU cache coherence [29, 30, 31, 41] are beneficial to our design and becoming common, they are not strictly necessary. Instead, CPU-GPU communication can also be achieved via atomic reads and writes in system memory and the GPU's device memory [28].

## 4 System Call Design Space Exploration

Enabling GPU system calls opens up many ways in which they can be used to improve programmability and performance. The "right" approach depends on the interactions between the algorithms and individual system calls. We discuss several interactions, providing GPGPU programmers a blueprint on reasoning about when and how to use system calls.

### 4.1 GPU-Side Design Considerations

To use system calls effectively in GPGPU code, programmers must be aware of the following considerations:

**Invocation granularity:** Potentially the first and most important question is, how should system calls be aligned with the hierarchy of GPU execution groups? GPU hardware relies on singe-instruction multiple-data (SIMD) execution and can concurrently run thousands of threads. To keep such massive parallelism tractable, GPGPU programming languages like OpenCL™ [9] and CUDA [19] exposes hierarchical groups of concurrently-executing threads to programmers. The smallest granularity of execution is the GPU work-item (akin to a CPU thread). Several work-items (typically, 32-64) operate in lockstep in the unit of wavefronts, the smallest hardware-scheduled units of execution. Several of these wavefronts (e.g., 16) make up programmer-visible work-groups and execute on a single compute unit (CU) of the underlying GPU hardware. Work-items in a work-group can communicate amongst themselves using local CU caches and/or scratchpads. Finally, hundreds of these work-groups together comprise the GPU kernel (GPU program). The CPU dispatches work to a GPU at the granularity of a kernel. Each work-group in a kernel can execute independently. Further, it is possible to synchronize just the work-items within a single work-group [9, 14]. This allows us to avoid the cost of globally synchronizing across thousands of work-items in a kernel, which is often unnecessary in a GPU program. In light of this, should a GPU system call be invoked separately for each work-item, once for every work-group, or once for the entire GPU kernel? Determining the right system call invocation granularity requires an understanding of the system call's use for the particular task at hand.

Consider a GPU program that writes sorted integers to a single output file. One might expect the *write* system call, invoked per work-item, to be the intuitive approach. This can present correctness issues, however, since *write* is position-relative, and requires access to a globally updated file pointer. Without synchronizing the execution of *write* across work-items, the output file will be written in a non-deterministic, and hence, unsorted order.

There exist several ways of fixing this problem using different system call invocation granularities. In one approach, the programmer can use a memory location to temporarily buffer the sorted output. Once all work-items have updated this buffer, the GPGPU programmer can use a single *write* system call at the end of the kernel to write the contents of the buffer to the output file. This approach partly loses the benefits of the GPU's parallel resources, since the entire system call latency is exposed on the program's critical path and cannot be overlapped with the execution of any other work-items. A compromise may be to invoke a *write* system call per work-group, buffering the sorted output of the work-items until the per-work-group buffers are fully populated. Finally, yet another alternative may be to use *pwrite* system calls instead of *write* system calls. Since *pwrite* allows programmers to specify the absolute file position where the output is to be written, per-work-item invocations present no correctness problems.

Naturally, these approaches have different performance implications. While per-work-item invocations of *pwrite* result in a flood of system calls, potentially harming performance, overly-coarse kernel-grain invocations also restrict performance by wasting the GPU's parallel execution resources. Sec. 6 shows that these decisions can yield as much as a $1.75\times$ performance difference.

As such, the utility of a system call and its invocation granularity are context-dependent and must be assessed carefully by the programmer. Therefore, we believe that an effective GPU system call interface must allow the programmer to specify the desired invocation granularity for every supported system call.

**System call ordering semantics:** When programmers use traditional system calls on the CPU, they expect program instructions appearing before invocation to finish completely. They also expect that the CPU resumes program execution only after the system call returns. Enforcing such "strong" ordering in the GPU may be natural for work-item granularity invocations but is problematic for work-group and kernel-grain invocations. Consider, for example, strong ordering on per-work-group invocations. We would need to place a barrier before the system call to ensure that all work-items in the work-group are synchronized. Similarly, we would require a second barrier after system call return so that work-items remain strongly ordered. Since a work-group can easily consist of hundreds of work-items, such barriers are expensive. It is worse for kernel granularity invocation since a typical kernel consists of several such work-groups, totaling thousands of work-items.

Fortunately, strong ordering is not always necessary in GPU programs. We thus introduce the concept of "relaxed ordering" of system calls. Fundamental to this concept is the observation that in general, in a program, a system call play a role of primarily a producer or a consumer of information. For example, system calls like *write* and *send* generate or write data while those like *read* and *receive* consume or read data. For correct operation, system calls in producer role typically require that all work-items corresponding to the system call's invocation granularity finish executing all their program instructions before the system call. Producers, however, may not necessarily have to wait for the system call to finish. For example, consider the *write* system call and assume that the programmer invokes it at the work-group granularity. Program instructions after the *write* call *may not* necessarily depend upon the outcome of the *write* itself; therefore, other work-items in the group need not wait for the completion of the system call. This is in contrast with system calls in consumer role like *read*, which typically require system calls to return before any work-items can start executing any program instructions after invocation. Consumers, however, do not necessarily require all work-items of the work-group to finish executing all instructions before the invocation. The same observations apply to per-kernel system call invocations.

Overall, relaxed ordering obviates the need for costly barrier synchronizations *around* system calls. For system call invocations in purely producer role, it eliminates the barrier after the invocation, and for those in only consumer role, it eliminates the barrier before invocation. Our analysis in Sec. 6 shows that ensuing performance improvements can be significant, up to 27%.

Beyond performance, we have found that relaxed ordering is necessary for functional correctness for kernel granularity invocation. This is because GPU kernels can consist of more work-items than can concurrently execute on the GPU. For strongly ordered system calls invoked at the kernel-grain, all work-items in the kernel need to finish executing instructions before the invocation but all work-items cannot execute concurrently. To make space, *some* work-items need to *completely* finish their execution, including system call invocation. Unfortunately, this contradicts strong ordering requirements. In other words, using strong ordering at kernel granularity can run the risk of deadlocking the GPU. We need relaxed ordering for these cases to work correctly.

In summary, at work-item granularity, we enable only implicit strong ordering (akin to CPU threads). At work-group granularity invocation, a programmer can choose between strong or weak ordering to fine tune trade-offs between programmability and performance. Finally, at kernel granularity only weak ordering is possible as forward progress may otherwise not be guaranteed.

**Blocking versus non-blocking approaches:** Most traditional CPU system calls, barring those for asynchronous I/O (e.g., *aio_read*, *aio_write*), return only after the system call finishes execution and are hence blocking. We find, however, that blocking system calls are often overly-restrictive for GPUs. In particular, because GPUs use SIMD execution, if even a single work-item is blocked on a system call, none of the other work-items in the wavefront can make progress either. The natural way of countering the overheads of system calls is to leverage the GPU's parallel execution model. That is, we find that GPUs can often benefit from *non-blocking* system calls which can return immediately, before system call processing completes. Non-blocking system calls can therefore overlap system call processing on the CPU with useful GPU work, improving performance by as much as 30% in some of our studies (see Sec. 6).

The concepts of blocking versus non-blocking and strong versus relaxed ordering are related but are orthogonal. The strictness of ordering refers to the question of when a system call can be invoked, with respect to the progress of work-items within its granularity (execution group) of invocation. Instead, the question of system call blocking refers to how the return from a system call relates to the completion of its processing. Therefore, they can be combined in several useful ways. For example, consider a case where a GPU program writes to a file at work-group granularity. The execution may not depend upon the output of the write but the programmer may want to ensure that write completes successfully. In such a scenario, blocking *writes* may be invoked with weak ordering. Weak ordering permits all but one wavefront in the work-group to proceed without waiting for completion of the *write*. This enables us to leverage the GPU's parallelism (see Section 5). Blocking invocation, however, ensures that one wavefront waits for the *write* to complete and can raise an alarm if the write fails. Consider another scenario, where a programmer wishes to prefetch data using *read* system calls but may not use the results immediately. Here, weak ordering with non-blocking invocation is likely to provide the best performance without breaking the program's semantics. In short, different combinations of blocking and ordering enables a GPU programmer to fine tune performance and programmability tradeoffs.

## 4.2 CPU-Side Design Considerations

Once GPUs interrupt the CPU to convey the need to process system calls, several design considerations determine the efficiency of CPU execution.

**System call coalescing:** GPUs rely on extreme parallelism to enable good performance. This means there may be bursts of system calls invoked on the CPU by the GPU. System call coalescing is one way to increase the throughput of system call processing. The idea is to collect several GPU system call requests and batch their processing on the CPU. The benefit is that CPUs can manage multiple system calls as a single unit, scheduling a single software task to process them. This reduction in task management, scheduling, and processing overheads can often boost performance (see Sec. 6). Coalescing must be performed judiciously as it improves system call processing throughput at the expense of latency. It implicitly serializes the processing of system calls within a coalesced bundle.

To allow the GPGPU programmer to balance the benefits of coalescing with its potential problems, we enable our system call interface to accept two parameterized bits of information – the time window length that the CPU waits for, in order to coalesce subsequent system calls, and the maximum number of system call invocations that can be coalesced within the time window. Our analysis in Sec. 6 reveals that system call coalescing can improve performance by as much as 10-15%.

# 5 Implementing GENESYS

We implemented GENESYS on a CPU-GPU system with the architectural details presented in Table 1. Our target system uses an AMD FX-9800P heterogeneous processor with an integrated GPU, and runs the open-source ROCm software stack [3]. We modified the GPU driver and Linux kernel code to enable GPU system calls. We also modified the HCC compiler in order to permit GPU system call invocations in its C++AMP dialect.

By construction, GENESYS supports the invocation of any system call from GPU programs. To enable a more tractable study, however, we focus on 11 system calls that span memory allocation as well as filesystem and network I/O.

Table 1: System Configuration

| | |
|---|---|
| **SoC** | Mobile AMD FX-9800P$^{TM}$ |
| **CPU** | 4× 2.7GHz<br>AMD Family 21h Model 101h |
| **Integrated-GPU** | 758 MHz AMD GCN 3 GPU |
| **Memory** | 16 GB Dual-Channel DDR4-1066MHz |
| **Operating system** | Fedora 25 using<br>ROCm stack 1.2<br>(based on Linux 4.4) |
| **Compiler** | HCC-0.10.16433<br>C++AMP with HC extensions |

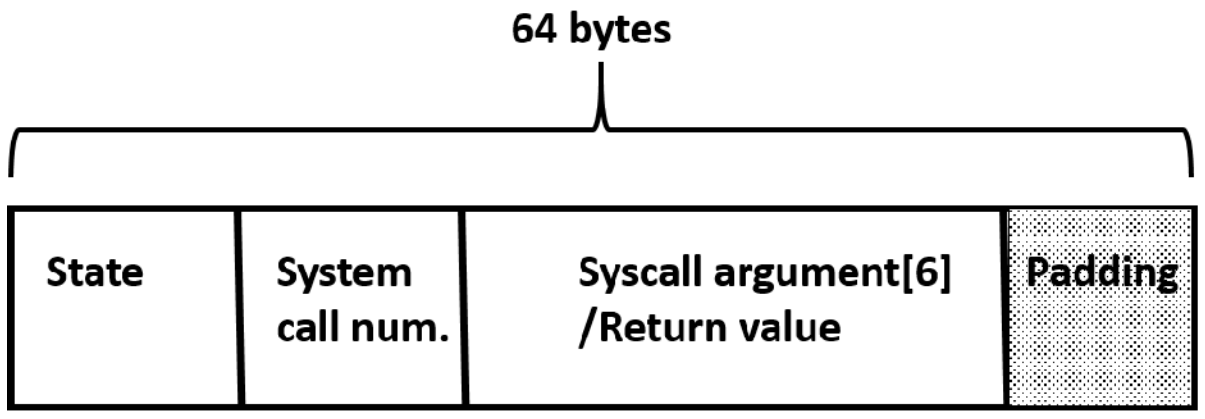

Figure 3: Content of each slot in syscall area.

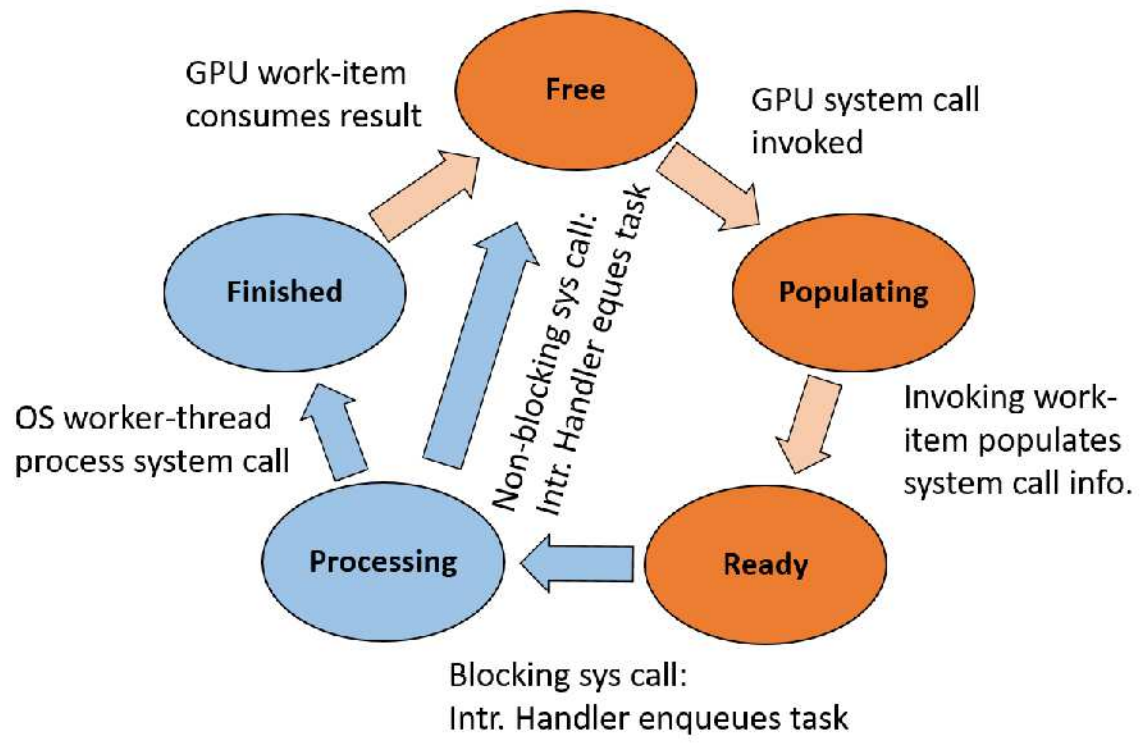

Figure 4: State transition diagram for a slot in syscall area. Orange shows GPU side state and actions while blue shows that of the CPU.

We designed GENESYS with several goals in mind. We desired a general framework capable of invoking any system call from the GPU program with the traditional function-like semantics used for CPUs. We also desired a simple implementation to enable easy adaptation to different hardware. Finally, prior work has shown that without care, GPU execution can hamper the CPU's access to shared resources like memory bandwidth and chip- or package-wide power and thermal budgets [27]. Hence, GENESYS also aims to limit interference with the CPU. To achieve these goals, we implemented the system with following high level parts:

**Invoking GPU system calls:** GENESYS permits programmers to invoke GPU system calls at the work-item-level. When the programmer intends to invoke system calls at a coarser-granularity (e.g., the work-group or kernel), a single work-item is designated as the caller on behalf of the entire work-group or kernel. System calls invoked at the work-item granularity follow CPU-centric strong ordering implicitly. However, work-group invocation granularities can be either relaxed or strongly ordered. With strong ordering, the GPU system call uses a work-group scope synchronization before and after system call invocation. When relaxed ordering is employed, a synchronization is placed either before (for system calls in producer role) or after (for system calls in consumer role) invoking the system call. Strong ordering at kernel invocations are unsupported as they may deadlock GPU hardware.

**GPU to CPU communication:** After the GPU invokes a system call, GENESYS facilitates efficient GPU to CPU communication of the system call request. As described in Section 3, GENESYS uses a pre-allocated shared memory location called a *syscall area* to allow the GPU to convey system call parameters to the CPU. The *syscall area* maintains one *slot* for each active GPU work-item. The OS and driver code can identify the desired slot by using the hardware ID of the active work-item, a piece of information available to modern GPU runtimes. Note that the hardware ID is distinct from the programmer-visible work-item ID. Each of potentially millions of work-items has a unique work-item ID that is visible to the application programmer. At any one point in time, however, only a subset of these work-items executes (as permitted by the GPU's CU count, supported work-groups per CU, and SIMD width). The hardware ID distinguishes amongst these active work-items. Overall, our system uses 64 bytes per slot, for a total of 1.25 MBs of total *syscall area* size.

Figure 3 depicts the information that is contained within each slot. The fields are the requested system call number, the request's state, system call arguments (as many as 6, in Linux), and padding to avoid false sharing. The field for arguments is also re-purposed for the return value of the system call. Figure 4 shows that when a GPU program's work- item invokes a system call, it atomically updates the state of the corresponding slot from *free* to *populating*. If the slot is not *free*, system call invocation is delayed. Once the state is *populated*, the invoking work-item populates the slot with system call information and changes the state to *ready*. The work-item also adds one bit of information about whether the invocation is blocking or non-blocking. The work- item then interrupts the CPU using a scalar GPU instruction (*s_sendmsg* on AMD GPUs). For blocking invocation, the work-item waits for the availability of the result by polling on the state of the slot or could suspend itself if the hardware supports. Note that, for a non-blocking request, the GPU does not wait for the result.

**CPU-side system call processing:** Once the GPU interrupts the CPU, system call processing commences. The interrupt handler creates a new kernel task and adds it to Linux's work-queue. This task is also passed the hardware ID of the requesting wavefront. At an expedient future point in time an OS worker thread executes this task. The task scans the *syscall area* of the given hardware ID and, atomically switches any *ready* system call requests to the *processing* state for blocking system calls. The task then carries out the system call work.

A subtle but critical challenge is that Linux's traditional system call routines implicitly assume that they are to be executed within the context of the original process invoking the system call. Consequently, they use context- specific variables (e.g., the *current* variable used to identify the process context). This presents a problem for GPU system calls, which are instead serviced purely in the context of the OS' worker thread. GENESYS overcomes this problem in two ways – it either switches to the context of the original CPU program that invoked the GPU kernel, or it provides additional context information in the code performing system call processing. The exact strategy is employed on a case-by-case basis.

GENESYS implements coalescing by waiting for a predetermined amount of time before enqueueing the task to process a system call, on an interrupt. If multiple requests are made to the CPU during this time window, they are coalesced with the first system call, such that they can be handled as a single unit of work by the OS worker-thread. GENESYS exposes two parameters to control the amount of coalescing through Linux's *sysfs* interface – the length of time window, and the number of maximum system calls that can be coalesced together.

**Communicating results from the CPU to the GPU:** Once the CPU worker-thread finishes processing the system call, results are put in the field for arguments in the slot for blocking requests. Further, it also changes the state of the slot to *finished* for blocking system calls. For non-blocking invocations, the state is changed to *free*. The invoking GPU work-item is then re-scheduled (on hardware that supports wavefront suspension) or automatically restarts because it was polling on this state. Results presented in this paper are from hardware where polling within the GPU is necessary. The work-item can consume the result and continue execution.

# 6 Evaluation with Microbenchmarks

Our evaluation begins with performance implications of different ways in which programmers can invoke GPU system calls. We use microbenchmarks to quantify the trade-offs in the different approaches.

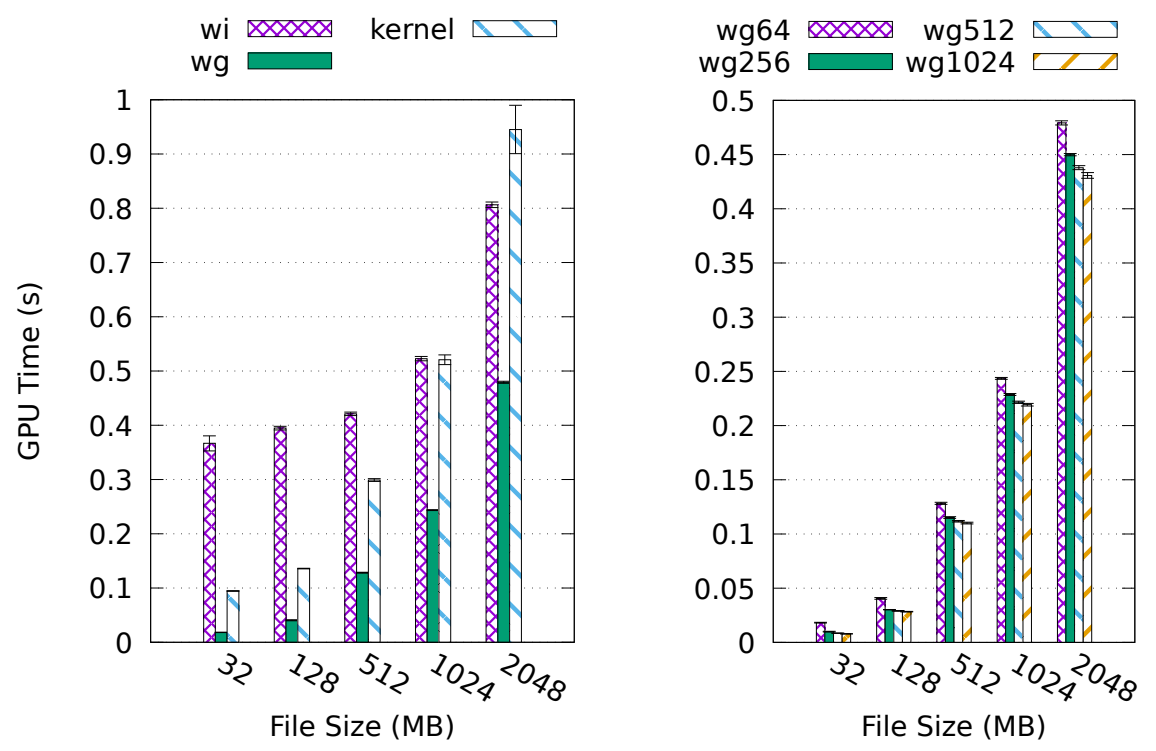

Figure 5: Impact of system call invocation granularity.

**Invocation granularity:** Whether a system call should be invoked at the work-item, work-group, or application kernel level is often application-dependent and is best left up to the programmer. The choice does have performance implications. Figure 5(a) quantifies this, for a GPU microbenchmark that uses *pread* to read data from files in *tmpfs*. The x-axis plots the file size, and y-axis shows the time to read the file, with lower values being better. Within each cluster, we separate runtimes for different *pread* invocation granularities.

Figure 5(left) shows that work-item invocation granularities tend to perform worst. This is unsurprising, since this is the finest granularity of system call invocation and leads to a flood of individual system calls that overwhelm the CPU. On the other end of the granularity spectrum, kernel-level invocation is problematic too as it generates a single system call and fails to leverage any potential parallelism in processing of system call requests. This problem is particularly severe at large file sizes (e.g., 2GB). In general, a good compromise is to use work-group invocation granularities instead. It does not choke the CPU cores with large bursts of system call requests while still being able to exploit parallelism in servicing system calls (Figure 5(left)).

In choosing how best to use the work-group invocation approach, an important design issue is the question of how many work-items should constitute a work-group. While Figure 5(left) assumes 64 work-items in a work-group, Figure 5(right) quantifies the performance impact of *pread* system calls as we vary work-group sizes from 64 to 1024 work-items. In general, larger work-group sizes enable better performance, as there are fewer unique system calls necessary to handle the same amount of work.

**Blocking and ordering strategies:** The choice of whether to use blocking or non-blocking system calls, with strong or weak ordering, also has a significant im-pact on performance. To quantify this, we designed a microbenchmark that uses the GPU to perform block permutation on an array. This block permutation is similar to the permutation steps performed in the popular DES encryption algorithm. The input data array is pre-loaded with random values and is divided into blocks of 8KB. We designed this microbenchmark so that work-groups execute 1024 work-items each and permute on blocks independently. At the end of the permutation, the results are written to a file by invoking the *pwrite* system call at work-group granularity. Importantly, it is possible to overlap *pwrite* system calls for one block of data with permutations on other blocks of data. To vary the amount of computation per system call, we repeat the permutation multiple times before writing the result.

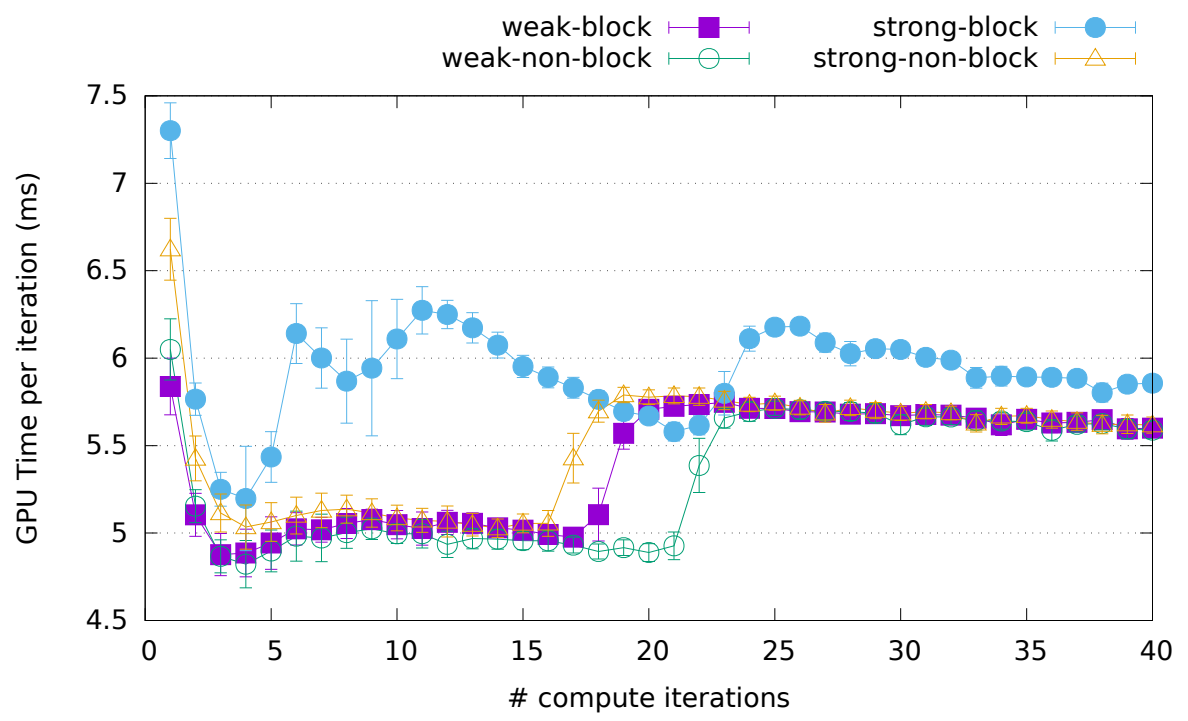

Figure 6: Performance implications of system call blocking and ordering semantics.

Figure 6 quantifies the impact of using blocking versus non-blocking system calls with strong and weak ordering. The trend lines separate results for the four combinations of strong- and weak-ordering with blocking and non-blocking system calls. The x-axis plots the number of permutation iterations performed on each block by each work-group before writing the results. Meanwhile, the y-axis plots the average time needed to execute one iteration (lower is better).

Figure 6 shows that strongly ordered blocking invocations (*strong-block*) hurt performance. This is expected as they require work-group scoped barriers to be placed before and after *pwrite* invocations. The GPU's hardware resources for work-groups are stalled waiting for the *pwrite* to complete. Not only does the inability to overlap GPU parallelism hurt strongly ordered blocking performance, it also means that GPU performance is heavily influenced by CPU-side performance vagaries like synchronization overheads, servicing other processes, etc. This is particularly true at iteration counts where system call overheads dominate GPU-side computation – below 15 compute iterations on the x-axis. Even when the application becomes GPU-compute bound, performance remains non-ideal.

Figure 6 shows that alternately, when *pwrite* is invoked in a non-blocking manner (albeit with strong ordering), performance is aided. This is expected since non-blocking *pwrites* permit the GPU to end the invoking work-group, freeing GPU resources it was occupying. The CPU-side processing of *pwrite* then can overlap with the execution of another work-group permuting on a different block of data utilizing just-freed GPU resources. Figure 6 shows that generally, latencies drop by 30% compared to blocking calls at low iteration counts. At higher iteration counts (beyond 16), these benefits diminish since the latency to perform repeated permutations dominates any system call processing times.

Next consider relaxing system call ordering but retaining blocking invocation (*weak-block*). In these cases, the work-group scope barrier post-system call is eliminated. One out of every 16 wavefronts [2] in the work-group waits for the blocking system call, while others exit, freeing GPU resources. Once again, the GPU can use these freed resources to run other work-items, hiding CPU-side system call latency. Consequently, the performance trends follow those of *strong-non-block*, with minor differences in performance arising from differences in the way the GPU schedules the work-groups for execution. Finally, Figure 6 shows that GPUs are most effectively able to hide system call latencies using relaxed and non-blocking approaches (*weak-non-block*).

**Interrupt coalescing:** Figure 7 shows the performance impact of coalescing multiple system calls. We again use a microbenchmark that invokes *pread* for our experiments. We read data from files of different sizes using a constant number of work-items. More data is read per *pread* system call from the larger files. The x-axis shows the varying amounts of data read, and quantifies the latency per requested byte. We present two bars for each point on the x-axis, illustrating the average time needed to read one byte with the system call in the absence of coalescing and when up to eight system calls are coalesced. In general, coalescing is most beneficial when small amounts of data are read. Reading more data generally takes longer; the act of coalescing the ensuing longer-latency system calls induces greater serialization in their processing on the CPUs, degrading performance.

## 7 Case Studies

In this section we present three workloads that execute on the GPU and exercise different subsystems of the OS: storage, memory management, and network. In targeting either better programmability or performance, or both, we showcase the generality and utility of GENESYS.

[2] Each wavefront has 64 work-items. Thus a 1024 work-item wide work-group has 16 wavefronts.

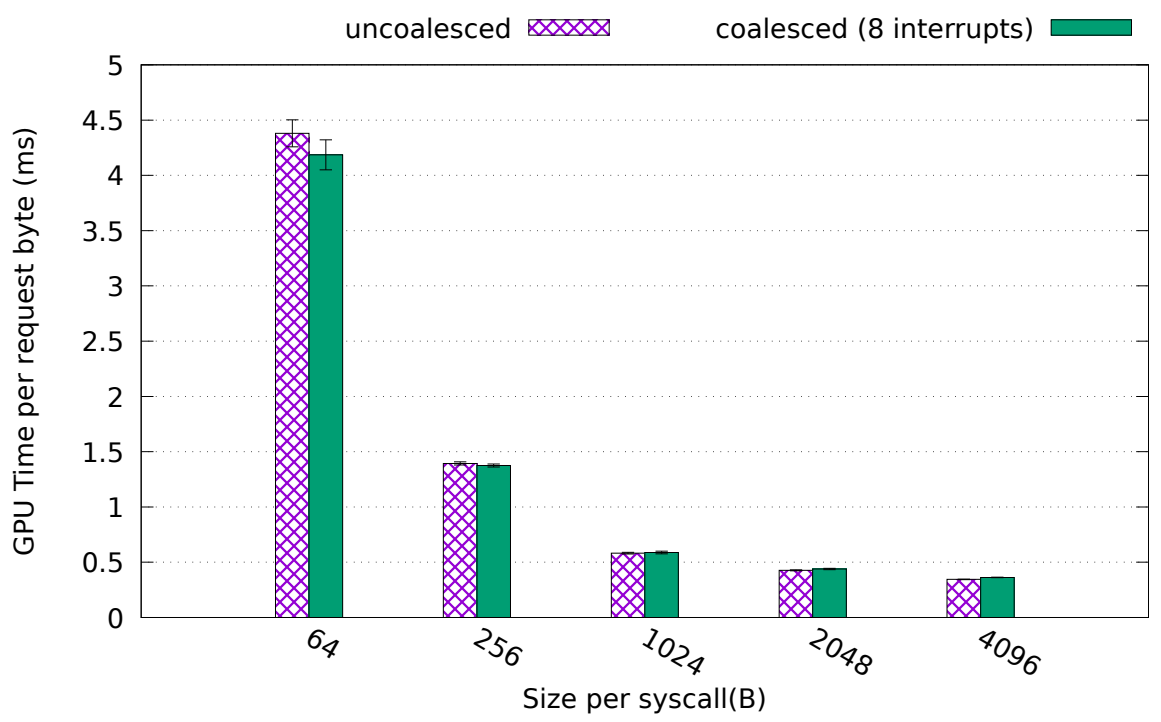

Figure 7: Implications of system call coalescing.

## 7.1 Storage

We first demonstrate the value of GPU system calls for storage I/O. Like past work [37], we wrote a wordcount workload which takes a list of words and list of files to search. Our workload searches for the specified words in those files, and generates the word counts as output. To showcase the utility of GPU system calls relative to CPU performance, we show results for two versions of this workload – a CPU-only workload and a GPU workload. We parallelize the CPU version with OpenMP, assigning files to cores. Our GPU version directly invokes the *open*, *read*, and *close* system calls made available with GENESYS. We found, due to the nature of the GPU algorithm suitable for wordcount, that these system calls were best invoked at work-group granularity with blocking and weak-ordering semantics. Finally, both CPU and GPU workloads are configured to search for occurrences of 64 strings in the Linux source.

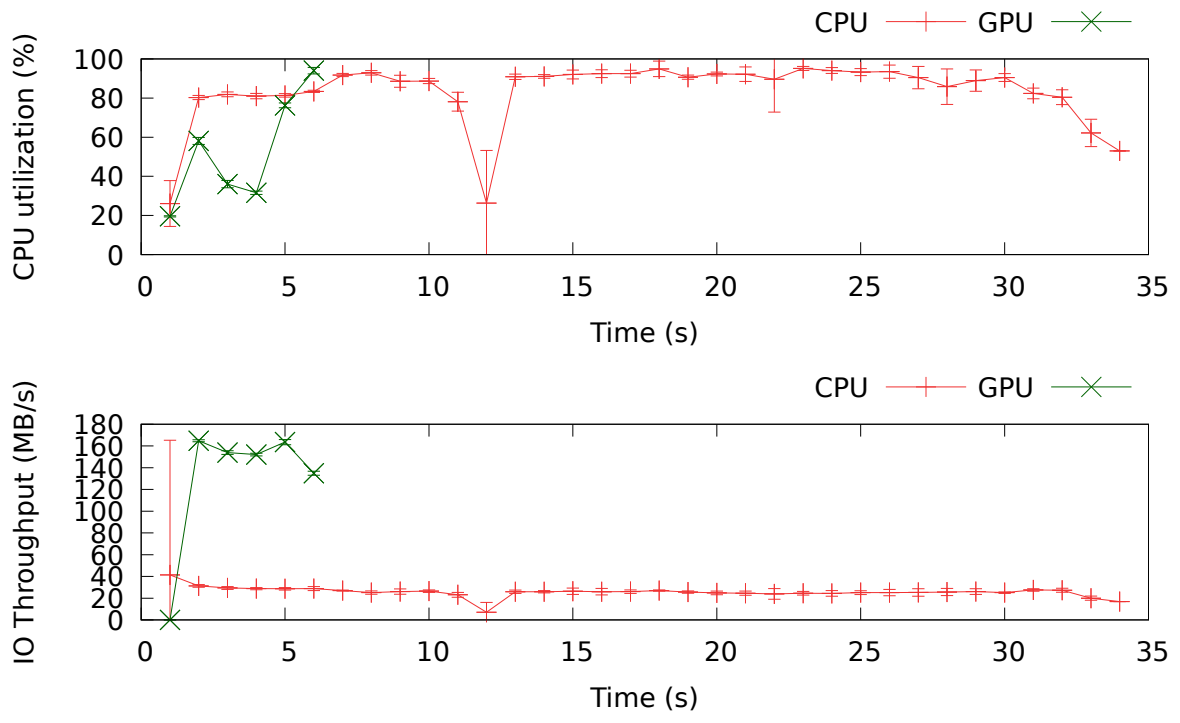

Figure 8: Wordcount I/O and CPU utilization reading data from an SSD. Note the high CPU usage necessary to provide high I/O throughput.

As one might expect, we found it much easier to write GPU code given access to direct system calls. But perhaps even more crucially, we found that the GPU version with direct system call invocation capabilities executes in roughly 6 seconds, constituting nearly a 6× performance improvement over the CPU version, which takes 35 seconds to complete. Figure 8 enables us to focus on this sharp performance improvement. We plot traces for CPU and GPU utilization and disk I/O throughput. Our results show that the GPU implementation extracts much higher throughput from the underlying storage device (up to 170MB/s compared to the CPU's 30MB/s). Offloading search string processing to the GPU frees up the CPU to process system calls effectively. The change in CPU utilization between the GPU workload and CPU workload reveals this trend. In addition, we made another interesting discovery – the GPU's ability to launch more concurrent I/O requests enabled the I/O scheduler to make better scheduling decisions, enhancing I/O throughput. This is made possible because of the ability to directly invoke many concurrent I/O requests from the GPU.

## 7.2 Memory

We next showcase the benefits of being able to use memory management system calls directly from the GPU. We focus our attention on the miniAMR application [34], and use the *madvise* system call directly from the GPU to better manage memory. The miniAMR application performs 3D stencil calculations using adaptive mesh refinement. This application is ripe for memory management because it varies its memory needs depending on its intended use. For example, when simulating regions experiencing turbulence, miniAMR needs to model with higher resolution. Instead, if lower resolution is possible without compromising simulation accuracy, miniAMR reduces memory and computation usage. As such, when miniAMR decides to reduce resolution, it is possible to free memory. While relinquishing excess memory in this way is not possible in traditional GPU programs, without explicitly dividing the offload into multiple kernels interspersed with CPU code (see 1), our framework does permit this with direct GPU *madvise* invocations. We invoke *madvise* using work-group granularities, with non-blocking and weak ordering.

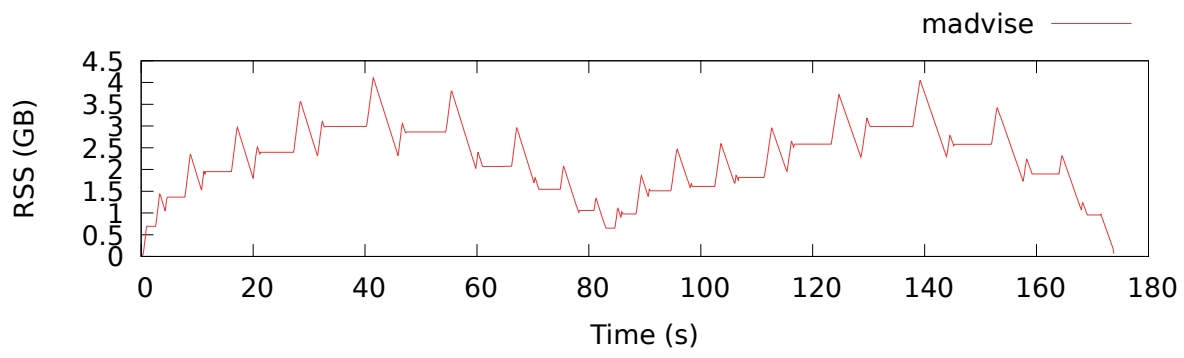

Figure 9: Memory footprint of miniAMR using madvise to hint at unused memory.

Figure 9 shows the results of our experiment. We execute miniAMR with an input dataset of 4.3GB. Figure 9 depicts a time trace of memory usage for miniAMR using *madvise*. The step-like nature of the curve is due to the periodic nature in which the memory reclamation process is executed. Note that the *madvise* additions to miniAMR help it use an RSS far below the peak usage of 4.3GB. As such, enabling such adaptivity to memory needs allows GPU programmers to relinquish memory, which is particularly valuable for workloads with massive amounts of data that often exceed the machine limits of physical memory capacity.

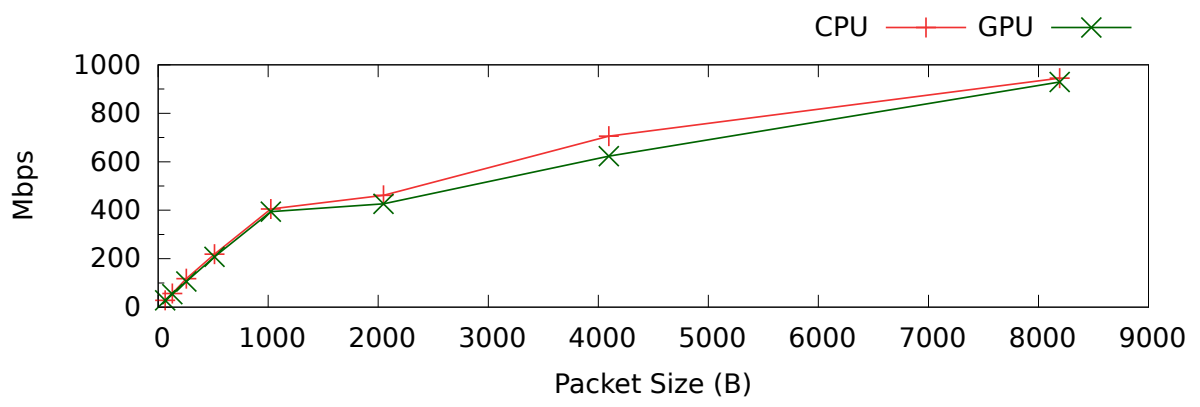

Figure 10: Bandwidth versus packet size for an echo server.

## 7.3 Network

Finally, we have also studied the programmability benefits of GPU system call support for networking. We wrote an echo server application that can process packets on either the GPU or CPU. Both the GPU and CPU portions of the application use *sendto* and *recvfrom* system calls. A load generator, running on the CPU, can choose to direct UDP packets to sockets on the CPU or to those running on the GPU. The GPU portion invokes these system calls at the granularity of a work-group and with blocking and weak ordering. Figure 10 depicts performance of when packets are directed to the CPU and when directed to the GPU. The x-axis is the packet size, while the y-axis is network throughput. Note the performance of the CPU and GPU are similar. Both can maximize link bandwidth with large packets. Both have similar trends.

# 8 Discussion

Enabling GPUs to invoke the full stack of generic system calls opens up several interesting research questions.

## 8.1 Linux system calls on GPUs

In the process of designing GENESYS, we considered all of Linux's roughly 300 system calls and assessed which ones to support. Figure 11 shows how we classify the system calls into several groups, explained below:

① **Useful and implementable:** This subset is useful and adaptable to GPUs today (left, lower corner in Figure 11). Examples include *pread, pwrite, mmap, munmap* etc. This is also the largest subset of Linux's system calls, comprising nearly 79% of them. In GENESYS, we implemented 11 such system calls for filesystems (*read, write, pread, pwrite, open, close*), networking (*sendto, recvfrom*), and memory management (*mmap, munmap, madvise*). Some of these system calls, like *read, write, lseek*, are stateful. Thus, their semantics may need to be extended for concurrent invocations for the same file descriptor. For example, the current value of the file pointer determines what value is read or written by the *read* or *write* system call. This can be arbitrary if invoked at work-item or work-group granularity for the same file descriptor since many work-items and work-groups are likely to execute concurrently.

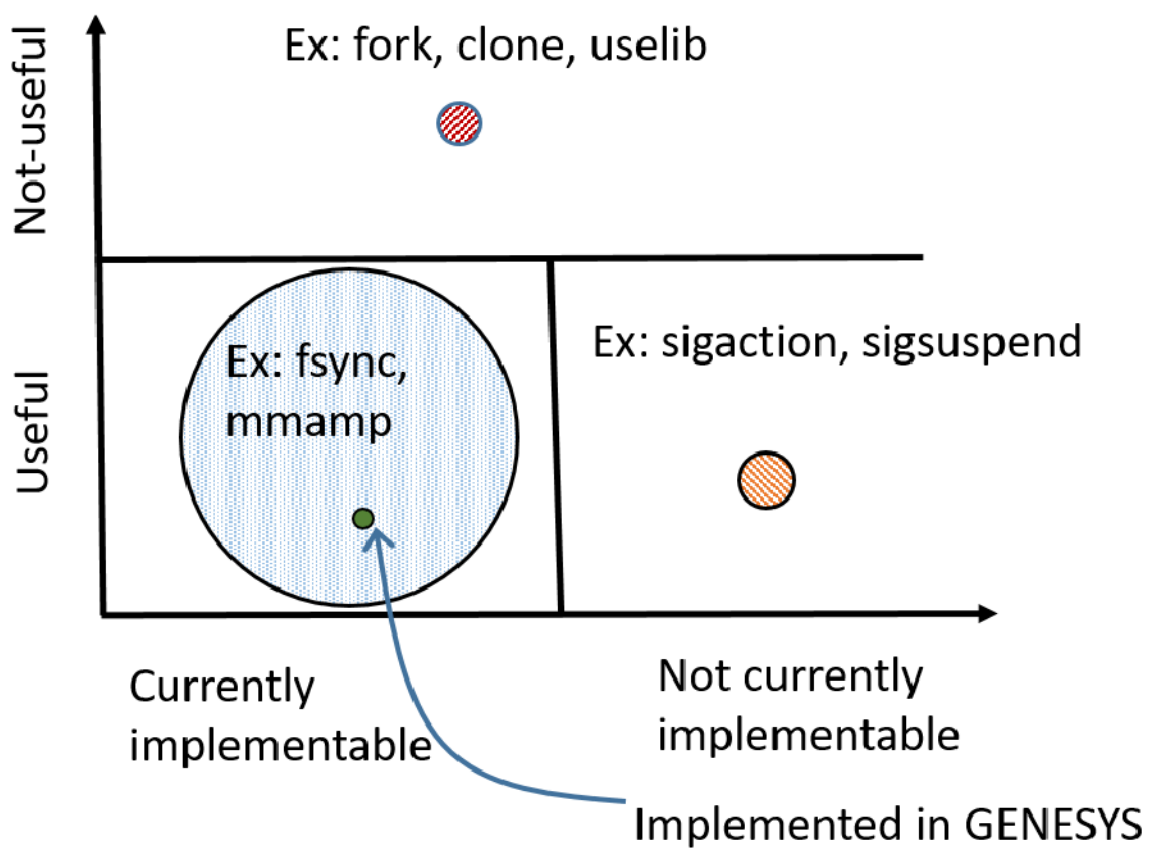

Figure 11: Our classification of Linux's system calls for GPUs. Size of each circle is representative of relative fraction of system calls in that category.

② **Useful but currently not implementable:** There are several system calls (13% of the total) that seem useful for GPU code, but are not easily implementable because of Linux's (POSIX's) existing design. For example, consider *sigsuspend* or *sigaction*. There is no kernel representation of a GPU work-item to manage and dispatch a signal to. Additionally, there is no lightweight method to alter the GPU program counter of a work-item from the CPU kernel. One approach is for signal masks to be associated with the GPU context and for signals to be delivered as additional work-items. These semantics do not match POSIX, but they seem useful.

③ **Not useful for GPUs at this time:** This small group (8% of the total) contains perhaps the most controversial set of system calls, as we debate even amongst ourselves whether or not they should ever be supported. Examples include *fork*, *clone*, *execv*, etc. At this time, we do not believe that it is worth the implementation effort to support these system calls from the GPU. For example, *fork* necessitates cloning a copy of the executing caller's GPU state. Technically, this can be done – it is how GPGPU code is context switched with the GUI for example – but it seems unnecessary to work at this time.

## 8.2 Lack of OS visibility

In the course of implementing GPU system call support within the Linux kernel, we came across a common design challenge. When a CPU thread invokes the OS, that thread has a *representation* within the kernel. The kernel maintains a data-structure for each thread and relies on it to accomplish several common tasks (e.g., kernel resource use, permission checks, auditing, etc.). GPU tasks, however, have traditionally not been represented in the OS kernel. *We believe this should not change.* GPU threads are numerous and short lived. Creating and managing a kernel structure for each GPU thread would vastly slow down the system. Moreover, these structures would only be useful for GPU threads that actually invoke the OS, and would be wasted overhead for the rest.

Our approach, given these constraints, is to process system calls that do not rely on the task representation in whatever thread context the CPU happened to be in at the time, and switch CPU contexts if necessary (See Section 5). As more GPUs have system call capabilities and more system calls are implemented, this is an area that will require careful consideration by kernel developers.

## 8.3 Asynchronous system call handling

In our design, GPU system calls are enqueued as kernel work items and processed outside of the interrupt context. We do this because Linux is designed such that few operations can be processed in an interrupt handler. A flaw with this design, however, is it defers the system call processing to potentially *past* the end of the life-time of the GPU thread and potentially the process that created the GPU thread itself! It is an example of a more general problem with asynchronous system calls [4]. Our solution is to provide a new function call, invoked by the CPU, that ensures all GPU system calls have completed.

## 8.4 Modifying GPU runtimes

In developing GENESYS we pushed the GPU HW/SW stack into operating regions it was not designed for, namely with work-groups that executed for seconds to minutes instead of microseconds. We've focused our design efforts on AMD hardware and software, but tests on other platforms reveal similar design assumptions. One peculiar effect we discovered is on our test platform is if a workgroup is long-lived, power-management kicks in and throttles *up* the GPU and *down* the CPU. While such a design choice is entirely sensible for graphics workloads, for GPGPU programs it may not always be. System call support on GPUs enables far richer application programs, where it may be entirely sensible to code GPU kernels where workgroups execute for the lifetime of the program. Future GPU HW/SW systems will need to take this into account. To help drive this innovation, we intend to release GENESYS as open source, and thereby encourage wide-spread adoption of a uniform method for invoking the OS from the GPU.

# 9 Related Work

Prior work has looked at subsystem-specific OS services for GPUs. GPUfs [37] provides filesystem access, while GPUNet [13] provides a socket-like API for network access. The latest generation of C++AMP [15] and CUDA [19] provide access to the memory allocator. These efforts employ a user-mode service thread executing on the CPU to proxy requests from the GPU [26]. As with GENESYS, system call requests are placed in a shared queue by the GPU. From there, however, the designs are different. The user-mode thread polls this shared queue and "thunks" the request to the libc runtime or OS. This incurs added overhead for entering and leaving the kernel on each request, instead of being naturally aggregated as with GENESYS. GPU workitems are also forced to spin, as user-mode code cannot wakeup the GPU hardware workgroup.

Several efforts have focused on providing network access to GPU code [38, 21, 18, 13]. NVidia provides GPUDirect [18], which is used by several MPI libraries [42, 8, 11], allows the NIC to bypass main memory and communicate directly to memory on the GPU itself. GPUDirect does not provide a programming interface for GPU side code. The CPU must initiate communication with the NIC. In [21] the author exposed the memory-mapped control interface of the NIC to the GPU and thereby allowed the GPU to directly communicate with the NIC. This low-level interface, however, lacks the benefits a traditional operating system interface brings (e.g. protection, sockets, TCP, etc.).

# 10 Conclusions

We present the first study that considers generic system calls on GPUs, showing their programmability and performance benefits. Not only does our system call interface support any system call, we also study various aspects of GPU system call invocation, such as ordering and blocking issues, invocation granularities, and CPU-side coalescing. Although more work needs to be done to enable truly heterogeneous CPU-GPU applications, our work is an important first step in making the GPU a peer to CPUs today.

## 11 Acknowledgment

We thank the National Science Foundation, which partially supported this work through grants 1253700 and 1337147. Finally, we thank Guilherme Cox for his feedback.

# Appendix

## A Classification of Linux System calls

| System call | Viable |
|---|---|
| sys_accept | yes |
| sys_accept4 | yes |
| sys_access | yes |
| sys_acct | yes |
| sys_add_key | yes |
| sys_adjtimex | yes |
| sys_alarm | yes, limited use* |
| sys_arch_prctl | yes |
| sys_bind | yes |
| sys_bpf | yes |
| sys_brk | yes, limited use** |
| sys_capget | no, targets threads*** |
| sys_capset | no, targets threads*** |
| sys_chdir | yes |
| sys_chmod | yes |
| sys_chown | yes |
| sys_chroot | yes |
| sys_clock_adjtime | yes |
| sys_clock_getres | yes |
| sys_clock_gettime | yes |
| sys_clock_nanosleep | yes **** |
| sys_clock_settime | yes |
| sys_clone/ptregs | yes |
| sys_close | yes |
| sys_connect | yes |
| sys_copy_file_range | yes |
| sys_creat | yes |
| sys_delete_module | yes |
| sys_dup | yes |
| sys_dup2 | yes |
| sys_dup3 | yes |
| sys_epoll_create | yes |
| sys_epoll_create1 | yes |
| sys_epoll_ctl | yes |
| sys_epoll_pwait | yes* |
| sys_epoll_wait | yes |
| sys_eventfd | yes |

| System call | Viable |
|---|---|
| sys_eventfd2 | yes |
| sys_execveat/ptregs | yes, limited use** |
| sys_execve/ptregs | yes, limited use** |
| sys_exit | yes***** |
| sys_exit_group | yes |
| sys_faccessat | yes |
| sys_fadvise64 | yes |
| sys_fallocate | yes |
| sys_fanotify_init | yes |
| sys_fanotify_mark | yes |
| sys_fchdir | yes |
| sys_fchmod | yes |
| sys_fchmodat | yes |
| sys_fchown | yes |
| sys_fchownat | yes |
| sys_fcntl | yes |
| sys_fdatasync | yes |
| sys_fgetxattr | yes |
| sys_finit_module | yes |
| sys_flistxattr | yes |
| sys_flock | yes, exclusive is limited** |
| sys_fork/ptregs | no |
| sys_fremovexattr | yes |
| sys_fsetxattr | yes |
| sys_fstatfs | yes |
| sys_fsync | yes |
| sys_ftruncate | yes |
| sys_futex | yes**** |
| sys_futimesat | yes |
| sys_getcpu | yes***** |
| sys_getcwd | yes |
| sys_getdents | yes |
| sys_getdents64 | yes |
| sys_getegid | yes |
| sys_geteuid | yes |
| sys_getgid | yes |
| sys_getgroups | yes |
| sys_getitimer | yes |
| sys_get_mempolicy | yes, address mode only |
| sys_getpeername | yes |
| sys_getpgid | yes |
| sys_getpgrp | yes |
| sys_getpid | yes |

| System call | Viable |
|---|---|
| sys_getppid | yes |
| sys_getpriority | yes***** |
| sys_getrandom | yes |
| sys_getresgid | yes |
| sys_getresuid | yes |
| sys_getrlimit | yes |
| sys_get_robust_list | no |
| sys_getrusage | yes, process level only |
| sys_getsid | yes |
| sys_getsockname | yes |
| sys_getsockopt | yes |
| sys_gettid | yes***** |
| sys_gettimeofday | yes |
| sys_getuid | yes |
| sys_getxattr | yes |
| sys_init_module | yes |
| sys_inotify_add_watch | yes |
| sys_inotify_init | yes |
| sys_inotify_init1 | yes |
| sys_inotify_rm_watch | yes |
| sys_io_cancel | yes |
| sys_ioctl | depends |
| sys_io_destroy | yes |
| sys_io_getevents | yes |
| sys_ioperm | no*** |
| sys_iopl/ptregs | yes |
| sys_ioprio_get | yes, CPU threads only |
| sys_ioprio_set | yes, CPU threads only |
| sys_io_setup | yes |
| sys_io_submit | yes |
| sys_kcmp | yes |
| sys_kexec_file_load | yes |
| sys_kexec_load | yes |
| sys_keyctl | yes |
| sys_kill | yes* |
| sys_lchown | yes |
| sys_lgetxattr | yes |
| sys_link | yes |
| sys_linkat | yes |
| sys_listen | yes |
| sys_listxattr | yes |
| sys_llistxattr | yes |
| sys_lookup_dcookie | yes |

| System call | Viable |
| --- | --- |
| sys_lremovexattr | yes |
| sys_lseek | yes |
| sys_lsetxattr | yes |
| sys_madvise | yes |
| sys_mbind | yes |
| sys_membarrier | no |
| sys_memfd_create | yes |
| sys_migrate_pages | yes |
| sys_mincore | yes |
| sys_mkdir | yes |
| sys_mkdirat | yes |
| sys_mknod | yes |
| sys_mknodat | yes |
| sys_mlock | yes |
| sys_mlock2 | yes |
| sys_mlockall | yes |
| sys_mmap | yes |
| sys_modify_ldt | yes |
| sys_mount | yes |
| sys_move_pages | yes |
| sys_mprotect | yes |
| sys_mq_getsetattr | yes |
| sys_mq_notify | yes* |
| sys_mq_open | yes |
| sys_mq_timedreceive | yes |
| sys_mq_timedsend | yes |
| sys_mq_unlink | yes |
| sys_mremap | yes |
| sys_msgctl | yes |
| sys_msgget | yes |
| sys_msgrcv | yes |
| sys_msgsnd | yes |
| sys_msync | yes |
| sys_munlock | yes |
| sys_munlockall | yes |
| sys_munmap | yes |
| sys_name_to_handle_at | yes |
| sys_nanosleep | yes**** |
| sys_newfstat | yes |
| sys_newfstatat | yes |
| sys_newlstat | yes |
| sys_newstat | yes |
| sys_newuname | yes |

| System call | Viable |
| --- | --- |
| sys_open | yes |
| sys_openat | yes |
| sys_open_by_handle_at | yes |
| sys_pause | no |
| sys_perf_event_open | yes, CPU perf events only |
| sys_personality | yes |
| sys_pipe | yes |
| sys_pipe2 | yes |
| sys_pivot_root | yes, limited use** |
| sys_pkey_alloc | yes |
| sys_pkey_free | yes |
| sys_pkey_get | yes |
| sys_pkey_mprotect | yes |
| sys_pkey_set | yes |
| sys_poll | yes |
| sys_ppoll | yes* |
| sys_prctl | yes |
| sys_pread64 | yes |
| sys_preadv | yes |
| sys_preadv2 | yes |
| sys_preadv64 | yes |
| sys_preadv64v2 | yes |
| sys_prlimit64 | yes |
| sys_process_vm_readv | yes |
| sys_process_vm_writev | yes |
| sys_pselect6 | yes* |
| sys_ptrace | yes** |
| sys_pwrite64 | yes |
| sys_pwritev | yes |
| sys_pwritev2 | yes |
| sys_pwritev64 | yes |
| sys_pwritev64v2 | yes |
| sys_quotactl | yes** |
| sys_read | yes |
| sys_readahead | yes |
| sys_readlink | yes |
| sys_readlinkat | yes |
| sys_readv | yes |
| sys_reboot | yes** |
| sys_recvfrom | yes |
| sys_recvmmsg | yes |
| sys_recvmsg | yes |
| sys_remap_file_pages | yes |

| System call | Viable |
|---|---|
| sys_removexattr | yes |
| sys_rename | yes |
| sys_renameat | yes |
| sys_renameat2 | yes |
| sys_request_key | yes |
| sys_restart_syscall | yes, no use* |
| sys_rmdir | yes |
| sys_rt_sigaction | yes* |
| sys_rt_sigpending | yes* |
| sys_rt_sigprocmask | yes* |
| sys_rt_sigqueueinfo | yes* |
| sys_rt_sigreturn/ptregs | yes, no use* |
| sys_rt_sigsuspend | yes, no use* |
| sys_rt_sigtimedwait | yes, no use* |
| sys_rt_tgsigqueueinfo | yes, no use* |
| sys_sched_getaffinity | yes, CPU threads only |
| sys_sched_getattr | yes, CPU threads only |
| sys_sched_getparam | yes, CPU threads only |
| sys_sched_get_priority_max | yes***** |
| sys_sched_get_priority_min | yes***** |
| sys_sched_getscheduler | yes, CPU threads only |
| sys_sched_rr_get_interval | yes, CPU threads only |
| sys_sched_setaffinity | yes, CPU threads only |
| sys_sched_setattr | yes, CPU threads only |
| sys_sched_setparam | yes, CPU threads only |
| sys_sched_setscheduler | yes, CPU threads only |
| sys_sched_yield | no |
| sys_seccomp | no |
| sys_select | yes |
| sys_semctl | yes |
| sys_semget | yes |
| sys_semop | yes |
| sys_semtimedop | yes |
| sys_sendfile64 | yes |
| sys_sendmmsg | yes |
| sys_sendmsg | yes |
| sys_sendto | yes |
| sys_setdomainname | yes** |
| sys_setfsgid | yes |
| sys_setfsuid | yes |
| sys_setgid | yes |
| sys_setgroups | yes |
| sys_sethostname | yes** |

| System call | Viable |
|---|---|
| sys_setitimer | yes * |
| sys_set_mempolicy | no |
| sys_setns | no |
| sys_setpgid | yes |
| sys_setpriority | yes***** |
| sys_setregid | yes |
| sys_setresgid | yes |
| sys_setresuid | yes |
| sys_setreuid | yes |
| sys_setrlimit | yes |
| sys_set_robust_list | no |
| sys_setsid | yes |
| sys_setsockopt | yes |
| sys_set_tid_address | no |
| sys_settimeofday | yes |
| sys_setuid | yes |
| sys_setxattr | yes |
| sys_shmat | yes |
| sys_shmctl | yes |
| sys_shmdt | yes |
| sys_shmget | yes |
| sys_shutdown | yes** |
| sys_sigaltstack | no |
| sys_signalfd | yes |
| sys_signalfd4 | yes |
| sys_socket | yes |
| sys_socketpair | yes |
| sys_splice | yes |
| sys_statfs | yes |
| sys_swapoff | yes** |
| sys_swapon | yes** |
| sys_symlink | yes |
| sys_symlinkat | yes |
| sys_sync | yes** |
| sys_sync_file_range | yes |
| sys_syncfs | yes** |
| sys_sysctl | yes** |
| sys_sysfs | yes** |
| sys_sysinfo | yes |
| sys_syslog | yes** |
| sys_tee | yes |
| sys_tgkill | yes* |
| sys_time | yes |

| System call | Viable |
|---|---|
| sys_timer_create | yes* |
| sys_timer_delete | yes |
| sys_timer_getoverrun | yes |
| sys_timer_gettime | yes |
| sys_timer_settime | yes |
| sys_timerfd_create | yes |
| sys_timerfd_gettime | yes |
| sys_timerfd_settime | yes |
| sys_times | yes, CPU times only |
| sys_tkill | yes* |
| sys_truncate | yes |
| sys_umask | yes |
| sys_umount | yes** |
| sys_unlink | yes |
| sys_unlinkat | yes |
| sys_unshare | yes |
| sys_userfaultfd | yes |
| sys_ustat | yes |
| sys_utime | yes |
| sys_utimensat | yes |
| sys_utimes | yes |
| sys_vfork/ptregs | no |
| sys_vhangup | yes |
| sys_vmsplice | yes |
| sys_wait4 | yes |
| sys_waitid | yes |
| sys_write | yes |
| sys_writev | yes |

Table 4: x86_64 system call table

* signals can be delivered only to CPU threads
** mostly serializing use, little benefit for GPU workloads
*** targets threads. There is currently no OS kernel structure representing thread level context for GPU task.
**** even though GPU threads are not represented, postponing return from system call will have the desired effect
***** implementable without system call, GPU modified semantics (local priority, CU instead of CPU core, ...)